\documentclass[iop]{emulateapj}
\usepackage{times}
\usepackage[breaklinks,colorlinks,urlcolor=blue,citecolor=blue,linkcolor=blue]{hyperref}
\usepackage{amsmath}
\journalinfo{The Astrophysical Journal, 2016, in press}

\shortauthors{Schiff \& Cranmer}
\shorttitle{Inverted Temperature Coronal Loops}

\begin{document}

\title{Explaining Inverted Temperature Loops in the Quiet
Solar Corona with Magnetohydrodynamic Wave Mode Conversion}

\author{Avery J. Schiff and Steven R. Cranmer}
\affil{Department of Astrophysical and Planetary Sciences,
Laboratory for Atmospheric and Space Physics,
University of Colorado, Boulder, CO 80309, USA}

\begin{abstract}
Coronal loops trace out bipolar, arch-like magnetic fields above the
Sun's surface.
Recent measurements that combine rotational tomography, extreme ultraviolet
imaging, and potential-field extrapolation have shown the existence of
large loops with inverted temperature profiles; i.e., loops for
which the apex temperature is a local minimum, not a maximum.
These ``down loops'' appear to exist primarily in equatorial quiet
regions near solar minimum.
We simulate both these and the more prevalent large-scale ``up loops''
by modeling coronal heating as a time-steady superposition of:
(1) dissipation of incompressible Alfv\'{e}n-wave turbulence, and
(2) dissipation of compressive waves formed by mode conversion from
the initial population of Alfv\'{e}n waves.
We found that when a large percentage ($>$~99\%) of the Alfv\'{e}n waves
undergo this conversion, heating is greatly concentrated at the
footpoints and stable ``down loops'' are created.
In some cases we found loops with three maxima that are also
gravitationally stable.
Models that agree with the tomographic temperature data exhibit higher
gas pressures for ``down loops'' than for ``up loops,'' which
is consistent with observations.
These models also show a narrow range of Alfv\'{e}n wave
amplitudes: 3 to 6 km~s$^{-1}$ at the coronal base.
This is low in comparison to typical observed amplitudes of 20 to 30
km~s$^{-1}$ in bright X-ray loops.
However, the large-scale loops we model are believed to comprise a
weaker diffuse background that fills much of the volume of the corona.
By constraining the physics of loops that underlie quiescent streamers,
we hope to better understand the formation of the slow solar wind.
\end{abstract}


\section{Introduction}
\label{sec:intro}

Coronal loops are strands of closed magnetic field that fill much of
the Sun's outer atmosphere with hot plasma.
Detailed simulations of loops are pivotal to understanding the physical
processes responsible for coronal heating and the Sun's overall
magnetohydrodynamic (MHD) evolution.
Early theoretical work employed constant heating rates as a function of
distance along each loop \citep[e.g.,][]{RTV}.
Later classical coronal loop modeling introduced spatially varying
heating rates from various sources but focused on short loops, often in
active magnetic regions \citep{Cg78,CP80,Wa82,SP90,Or95,Sp03,Lq08,Ma10}.
Typically, these studies describe loops with temperatures that increase
steadily from the chromosphere ($T \approx 10^{4}$~K) to the corona
($T \approx 10^{6}$~K).
When longer loops were simulated with heating rates concentrated near
their basal footpoints \citep{Se81,AS02}, solutions were found with
{\em decreasing temperatures with increasing height} in the vicinity of
their apex.
These loops were sometimes believed to be gravitationally or thermally
unstable \citep{As01}, but they remained unobserved for some time.

Observations of loops in X-ray and ultraviolet wavelengths have tended
to focus on the brightest and most collimated structures that stand out
in contrast against a significantly dimmer diffuse coronal background.
However, the recent development of differential emission measure
tomography \citep[DEMT; see][]{Fz05,Fz09,Va10,Va11} opened new
perspectives on the measurement of plasma parameters in larger
coronal structures.
DEMT combines two techniques to achieve superior resolution: measurement
of the coronal differential emission measure (DEM) distribution in
three-dimensional space using solar rotational tomography (SRT).
Multiple narrow-band images from the Atmospheric Imaging Array
\citep[AIA;][]{Lm12} on the {\em Solar Dynamics Observatory} ({\em{SDO}})
were combined to determine the relative amounts of plasma at different
temperatures along the optically thin line-of-sight.
Tomographic techniques were then employed---assuming negligible coronal
evolution over the time the Sun took to rotate through the full field
of view---to extract the DEM information in each volume element.
By tracing a field line provided by extrapolation of the coronal
magnetic field from photospheric magnetograms and a Potential Field 
Source Surface (PFSS) model, the technique provides the mean electron 
density and electron temperature at various points in a tomographic grid.
It then becomes possible to map out variations of temperature and pressure 
along individual large-scale closed field lines.

\citet{Hu12} employed this technique to conduct one of the first major
surveys of quiet-Sun loops.
In the process, they discovered seemingly stable loops with negative
temperature gradients near their apexes, which they labeled
``down loops.''
{At nearly the same time, additional evidence for down loops
was found from coronagraph measurements of line intensity ratios by
\citet{KP13}.}
The DEMT results were expanded upon by \citet{Nu13}, who quantified the
properties of several thousand loops in order to distinguish the
down loops from the more common ``up loops.''

This paper attempts to explain the results of \citet{Nu13} by simulating
large grids of coronal loops with heating mechanisms of varying strengths.
We use time-steady rates of coronal heating in our models.
There is considerable evidence for highly dynamic and time-dependent
energy deposition in the corona
\citep[see, e.g.,][]{As06,PD12,Kl15,Fl15,Ha15}.
However, the large, long-lived loops that are traceable using DEMT
represent quasi-steady thermal states averaged over a relatively long
time (i.e., several days).
Thus, we believe it makes sense to consider similarly time-averaged heating
rates in order to ascertain how that quasi-steady state is maintained.
We make use of recent successes in modeling coronal heating via the
nonlinear cascade of MHD turbulence, and we include the transfer of
energy between incompressible (Alfv\'{e}n) and compressible
(magnetoacoustic) fluctuations.

In Section \ref{sec:field} we describe an empirically motivated
way of modeling the shapes and magnetic properties of the large loops
observed by \citet{Nu13}.
Section \ref{sec:conserv} gives the relevant conservation equations
for time-steady loops, and
Section \ref{sec:heat} discusses our adopted prescriptions for
coronal heating.
In Section \ref{sec:method} we describe the numerical methods used to
solve for time-steady loop properties, the relevant boundary conditions,
and the search for unstable solutions.
Results for a large grid of loops are presented in
Section \ref{sec:results}, and we compare our models to the relevant
observations.
Finally, Section \ref{sec:conc} contains a summary of our results and
a discussion of implications for the overall coronal heating problem.

\section{Loop Magnetic Fields}
\label{sec:field}

In many models of small coronal loops, the magnetic geometry is either
simplified as a constant cross-section tube or as an expanding bundle
of field lines with a specified function for the area expansion rate.
For the large and often trans-equatorial loops considered in this paper,
we aim to use a more realistic approach by modeling the spatial behavior
of the vector magnetic field ${\bf B}$.
Below we describe a semi-analytic ``submerged monopole'' model for coronal
loop magnetic fields that is used in the coronal heating models below.
We also describe how the parameters of this model are constrained by
comparison with PFSS extrapolations of similar coronal conditions as
observed by \citet{Nu13}.

When tracing along the magnetic field with a distance coordinate $s$
that follows the field lines, magnetic flux conservation constrains
the field strength $B(s)$ to be inversely proportional to the
cross-sectional area $A(s)$ of an idealized flux tube.
For quiet-Sun loops far from active regions, we assume that the field
can be approximated as the gradient of a potential.
This allows us to construct the magnetic filed from the superposition
of a number of point-like monopole sources beneath the solar surface.
As long as the flux from positive sources is balanced by the flux from
negative sources, the magnetic field above the surface will have
$\nabla \cdot {\bf B} = 0$.
Thus,
\begin{equation}
  {\bf B} ({\bf r}) \, = \, \sum_{i} \frac{\Phi_{i}}{2\pi}
  \frac{{\bf r} - {\bf r}_{i}}{| {\bf r} - {\bf r}_{i} |^3} \,\, ,
  \label{eq:Bdef}
\end{equation}
where the coordinates ${\bf r}_{i}$ specify the locations of each
monopole source $i$, and the field point ${\bf r}$ can be located
anywhere at or above the solar surface \citep[see, e.g.,][]{Wa98,Cl03}.
$\Phi_i$ is the signed magnetic flux in each source, and we require
\begin{equation}
  \sum_{i} \Phi_{i} \, = \, 0 \,\,\, .
\end{equation}
For simplicity, we use only two sources: one positive (located
north of the equator) and one negative (south of the equator).
Because our coronal heating model does not depend on the absolute
magnitude of $B$, we consider the source fluxes $\Phi_i$ to be
arbitrary and set them to $\pm 1$.

\begin{figure}
\epsscale{1.2}
\plotone{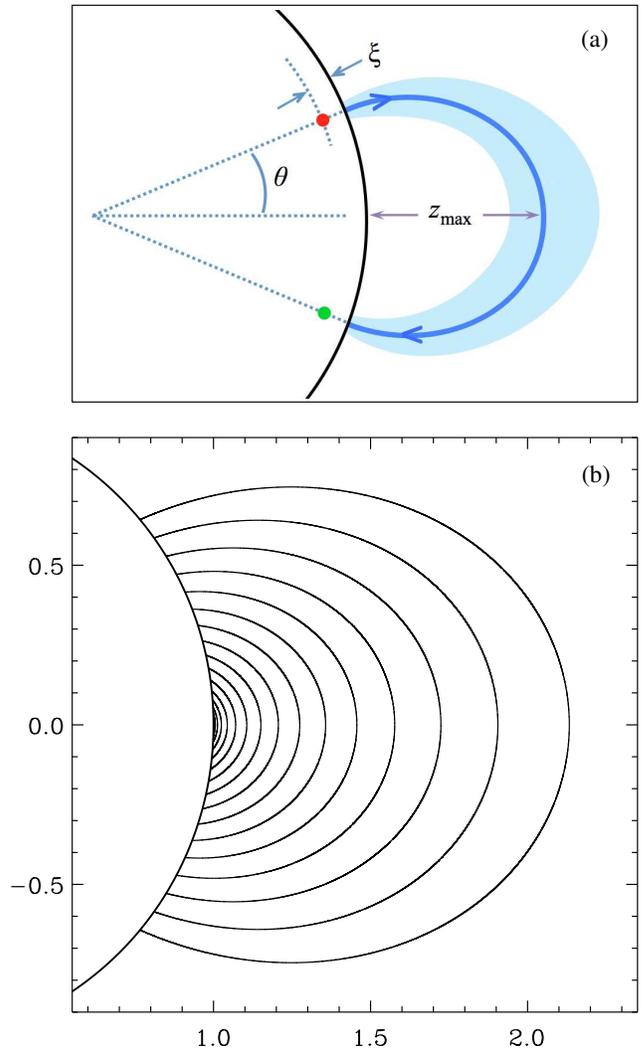}
\caption{Illustration of the adopted loop geometry.
(a) The loop length $L$ is measured from footpoint to footpoint
along the dark blue magnetic field line.
The monopole sources, submerged a distance $\xi$ beneath the surface,
are denoted by red and green circles.
(b) Loops computed with the $\xi(\theta)$ constraint of
Equation~(\ref{eq:bestxi}).
Increased values of $\theta$ lead to longer loops with greater
values for the maximum distance to the surface $z_{\rm max}$.
\label{fig01}}
\end{figure}

We constructed a grid of magnetic field properties for symmetric
loops characterized by two free parameters: a footpoint latitude
$\theta$ (i.e., half the angular separation between the two sources)
and the submerged depth $\xi$ of each source.
Figure~\ref{fig01}(a) illustrates the geometry and the two parameters.
Each choice of $\theta$ and $\xi$ describes a unique coronal loop
with length $L$ (measured from one footpoint to the other),
apex height $z_{\rm max}$ above the surface, and magnetic field
ratio $B_{\rm max}/B_{\rm min}$ (where $B_{\rm max}$ is found at
the surface and $B_{\rm min}$ is found at the apex).
The local vector field {\bf B} was traced along discrete steps of
length $ds = 1.5 \times 10^{-3} \, R_{\odot}$ and tabulated.
For symmetric fields, it is only necessary to trace the loop from
one footpoint ($s=0$) to its apex ($s = L/2$).

The full two-dimensional parameter space ($\theta$, $\xi$) of
possible loops encompasses many unrealistic cases.
In order to construct loops relevant to the actual quiet-Sun regions
studied by \citet{Hu12} and \citet{Nu13}, we found a single
one-dimensional cut through the parameter space by constraining
$\xi$ to be a function of $\theta$.
We found this by comparing with coronal magnetic fields extrapolated
from measured photospheric magnetograms with the PFSS technique.
The PFSS method assumes the corona is current-free below a spherical
``source surface'' at $r = 2.5 \, R_{\odot}$, and that the field is
pointed radially above it \citep{AN69,Sh69}.
We used a low-resolution PFSS model made with synoptic magnetogram data
from the Wilcox Solar Observatory \citep[WSO; see][]{HS86}.
To best match the properties of the loops observed by
\citet{Nu13}, we obtained PFSS coefficients for Carrington
Rotations 2065, 2077, 2081, and 2106.
These coefficients were constructed with maximum order $\ell = 9$ in
the spherical harmonic expansion.
Following recent practice \citep[e.g.,][]{DR12}, we zeroed out all
residual monopole ($\ell=0$) terms.
For each rotation, we traced outwards from 500 random locations on
the solar surface and kept track of only the closed loops.
The trends observed in the loop geometries were consistent across
numerous randomized sets.

\begin{figure}
\epsscale{1.15}
\plotone{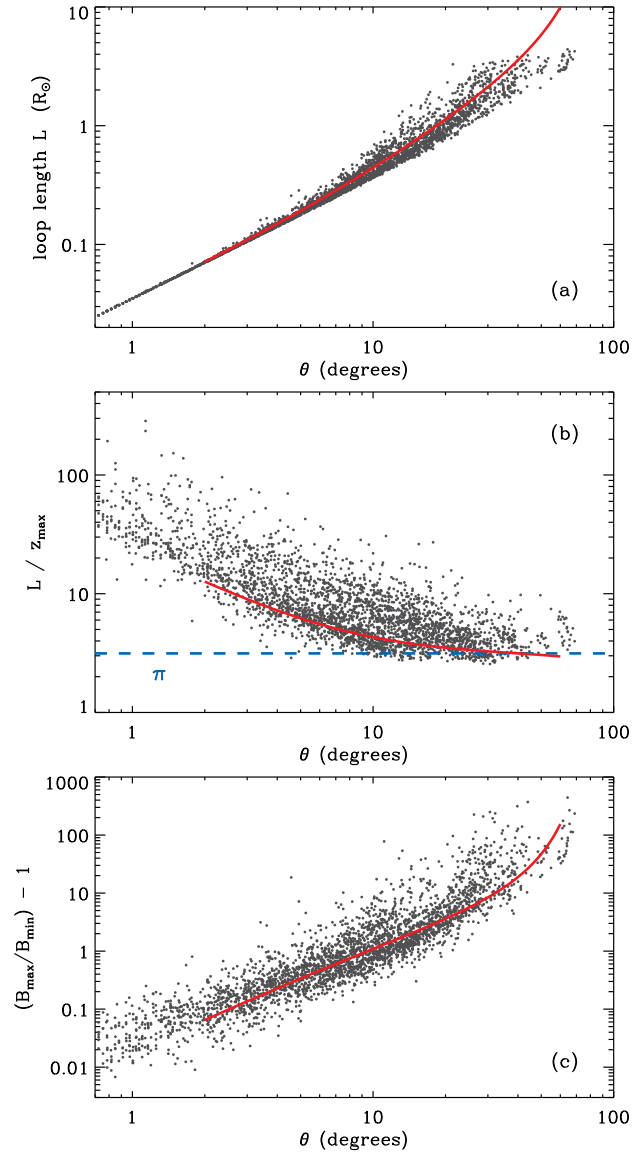}
\caption{(a) Model loop length $L$, (b) ratio of $L$ to the apex
height $z_{\rm max}$, and (c) ratio of maximum to minimum magnetic
field strength, all plotted versus footpoint latitude $\theta$.
Black symbols show PFSS model properties from Carrington Rotations
2065, 2077, 2081, and 2106.
Red curves show the result of applying Equation~(\ref{eq:bestxi}).
\label{fig02}}
\end{figure}

Figure~\ref{fig02} shows some representative parameters as a
function of $\theta$, which we measured from the PFSS model as half
of the great-circle angle between each loop's two footpoints.
We searched for best relationship between $\theta$ and $\xi$ in the
models described by Equation (\ref{eq:Bdef}), and we found
\begin{equation}
  \frac{\xi}{R_{\odot}} \, = \, 0.27 + \frac{2 \theta}{\pi}
  \label{eq:bestxi}
\end{equation}
where $\theta$ is expressed in radians.
Note that no single monotonic function $\xi (\theta)$ was able to
reproduce ``best-fitting'' curves for each of the trends shown in
Figure~\ref{fig02}; the above relation
is a compromise that appears to create reasonably realistic loops over
the full range of observed lengths.
Table~1 provides some basic data for a selection of loops that follows
this pattern, and Figure~\ref{fig01}(b) illustrates loop shapes for a
subset of parameters.
Note how the ratio $L/z_{\rm max}$ diverges from the idealized value of
$\pi$ that one would expect for a semi-circular loop.
The shortest loops are squat (i.e., $L/z_{\rm max} \gg \pi$)
and the longest loops are stretched out radially
($L/z_{\rm max} \lesssim \pi$).

\begin{deluxetable}{cccccc}
\tablecaption{Geometric Properties of Quiet Sun Loops
\label{table01}}
\tablewidth{0pt}
\tablehead{
\colhead{$\theta$} &
\colhead{$\xi$} &
\colhead{$z_{\rm max}$} &
\colhead{$L$} &
\colhead{$B_{\rm max}/B_{\rm min}$} \\
\colhead{(deg)} &
\colhead{($R_{\odot}$)} &
\colhead{($R_{\odot}$)} &
\colhead{($R_{\odot}$)} &
\colhead{---}
}
\startdata
   1  &  0.2811  &  0.0016  &  0.033  &  1.0187  \\
   2  &  0.2922  &  0.0059  &  0.069  &  1.0673  \\
   4  &  0.3144  &  0.0211  &  0.147  &  1.2328  \\
   8  &  0.3589  &  0.0711  &  0.333  &  1.7597  \\
  10  &  0.3811  &  0.1036  &  0.441  &  2.1018  \\
  12  &  0.4033  &  0.1402  &  0.561  &  2.4937  \\
  15  &  0.4367  &  0.2022  &  0.756  &  3.1797  \\
  20  &  0.4922  &  0.3242  &  1.134  &  4.6366  \\
  25  &  0.5478  &  0.4717  &  1.584  &  6.6300  \\
  30  &  0.6033  &  0.6499  &  2.121  &  9.4423  \\
  35  &  0.6589  &  0.8669  &  2.772  &  13.571  \\
  40  &  0.7144  &  1.1355  &  3.567  &  19.925  \\
  45  &  0.7700  &  1.4752  &  4.560  &  30.271  \\
\enddata
\end{deluxetable}

\begin{figure}
\epsscale{1.15}
\plotone{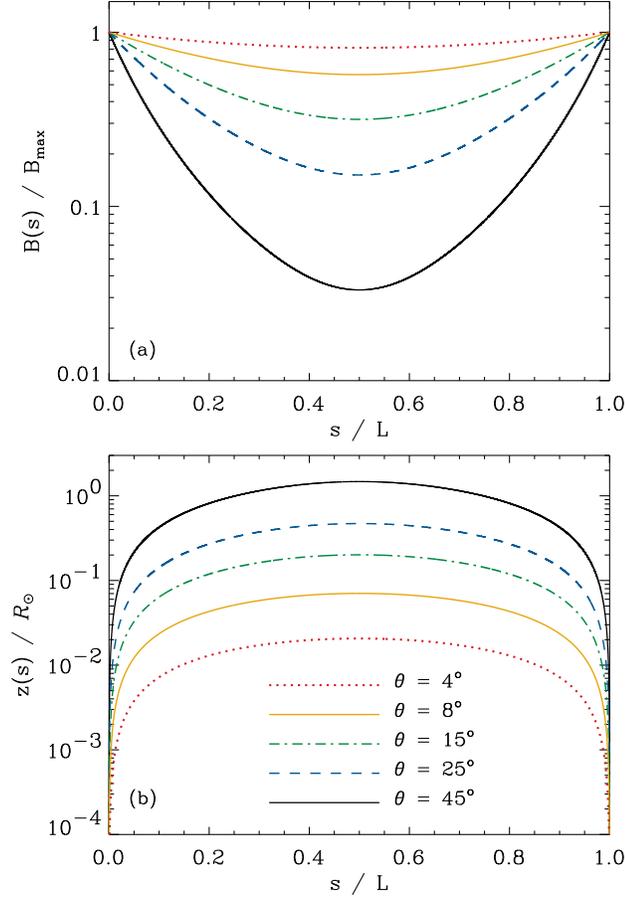}
\caption{Distance dependence of (a) magnetic field strength $B$,
in units of the maximum footpoint field strength $B_{\rm max}$,
and (b) height $z$ above the photosphere.
All properties are symmetric about the loop apex that occurs at
$s = L/2$.
\label{fig03}}
\end{figure}

Figure~\ref{fig03} illustrates the variation of magnetic field strength $B$
and height $z$ as a function of $s$ for loops of various lengths.
For a given loop configuration, the models need to know the relative
variation of $A(s) \propto B(s)^{-1}$ and also how the height $z$
(and heliocentric radius $r = z + R_{\odot}$)
depends on the loop-distance coordinate $s$.

\section{Conservation Equations}
\label{sec:conserv}

The plasma properties in the model coronal loops are computed
under the assumptions of time independence, complete ionization,
and hydrostatic equilibrium.
We do not consider the properties of the underlying photosphere
and chromosphere, and instead define the
surface ($s=0$, $z=0$) at the base of the transition region (TR),
with a fixed temperature $T_0$ and mass density $\rho_0$.
Section \ref{sec:method:bc} gives more details about the boundary
conditions.
We follow \citet{SvB05} to specify the equation of state as
\begin{equation}
  P = C_{1} n_{p} k_{\rm B} T
  \,\, , \,\,\,\,\,\,\,\,\,\,\,\,
  \rho = C_{2} n_{p} m_{p}
  \,\, , \,\,\,\,\,\,\,\,\,\,\,\,
  n_{e} = C_{3} n_{p}  \,\,\, ,
  \label{eq:eos}
\end{equation}
where $P$ is the gas pressure, $n_p$ is the proton density,
$n_e$ is the electron density, $k_{\rm B}$ is Boltzmann's constant,
and $m_p$ is the proton mass.
For a fully ionized hydrogen/helium plasma, the constants $C_i$ are
given by
\begin{equation}
  C_{1} = 2 + 3a
  \,\, , \,\,\,\,\,\,\,\,\,\,\,\,
  C_{2} = 1 + 4a
  \,\, , \,\,\,\,\,\,\,\,\,\,\,\,
  C_{3} = 1 + 2a
\end{equation}
and the helium abundance is given by $a \equiv n_{\rm He}/n_{\rm H}$.
We adopt $a = 0.05$, which is lower than the photospheric value
due to a combination of gravitational settling, thermal forces,
and first ionization potential (FIP) fractionation effects
\citep[see, e.g.,][]{LF03,Ki05}.

In order to determine appropriate temperature profiles for the loops,
we iteratively solve two differential equations.
The first is the conservation of momentum, given for a steady-state
environment in hydrostatic equilibrium by
\begin{equation}
  \frac{\partial P}{\partial s} \, = \,
  - \rho \, \frac{\partial \Phi_g}{\partial s}
  \label{eq:hyeq}
\end{equation}
where the gravitational potential is given by
\begin{equation}
  \Phi_{g} \, = \, -\frac{GM_{\odot}}{r} \,\,\, .
\end{equation}
For a given loop geometry, the quantity $\partial \Phi_g / \partial s$
is a known function of $s$ and is tabulated.
Using a steady-state version of the pressure equation from \citet{SvB05},
the loop pressure profile is given by
\begin{equation}
  P(s) \, = \, P_{0} \, \exp \left[ - \int_{0}^{s}
  \frac{C_{2} m_{p}}{C_{1} k_{\rm B} T(s')} \,
  \frac{\partial \Phi_g}{\partial s'} \, ds' \right]
  \label{eq:f}
\end{equation}
which depends on the base pressure $P_0$ and the full solution
$T(s)$ for temperature dependence along the loop.

The second conservation equation we solve is the time-steady and
hydrostatic version of the thermal energy equation,
\begin{equation}
  Q_{\rm heat} + Q_{\rm rad} + Q_{\rm cond} \, = \, 0
  \label{eq:mainthermal}
\end{equation}
where each $Q$ term is a volumetric rate of heating or cooling
(i.e., power per unit volume).
The coronal heating term $Q_{\rm heat}$ is discussed at length in
Section~\ref{sec:heat}.
We use the conventional expression for radiative cooling in an
optically thin plasma,
\begin{equation}
  Q_{\rm rad} \, = \, - n_{e}^{2} \, \Lambda(T)
\end{equation}
where we use the radiative loss function $\Lambda(T)$ tabulated by
\citet{CvB07}.
The net rate of conductive heating (or cooling) can be expressed as
\begin{equation}
  Q_{\rm cond} \, = \, \frac{1}{A} \, \frac{\partial}{\partial s}
  \left[ A \, \kappa(T) \, \frac{\partial T}{\partial s} \right]
\end{equation}
where the thermal conductivity $\kappa$ is defined in the same way
as in \cite{SvB05},
\begin{equation}
  \kappa(T) \, = \, \kappa_{0} T^{2.5} + \kappa_{1} T^{-0.5}
  \,\, .
  \label{eq:thermcond}
\end{equation}
The first term in Equation~(\ref{eq:thermcond}) accounts for electron
thermal conduction \citep{S62}, where
$\kappa_{0} = 8 \times 10^{-7}$ erg s$^{-1}$ cm$^{-1}$ K$^{-3.5}$.
This assumes a value for the Coulomb logarithm of 23, which is consistent
with coronal plasma having $T \sim 10^{6}$~K and
$n \sim 10^{6}$~cm$^{-3}$.
The second term in Equation~(\ref{eq:thermcond}) was included by
\citet{SvB05}, based on models of \citet{Fo90}, in order to account
for ambipolar diffusion in the TR.  Thus, we use
$\kappa_{1} = 4 \times 10^{9}$ erg s$^{-1}$ cm$^{-1}$ K$^{-0.5}$,
which is set so that the total conductivity $\kappa$ is minimized at
$T = 10^{5}$~K.

\section{Coronal Heating from Wave Dissipation}
\label{sec:heat}

The fundamental physical processes that deposit heat into the corona
are still not known.
However, most agree on the broad outlines of a scenario in which
magnetic field lines at the surface are jostled by convective motions.
Some of that kinetic energy propagates up to larger heights and is
converted into magnetic energy.
The magnetic field is continually driven by these motions into a complex 
collection of small-scale, nonlinear distortions.
Lastly, once the spatial scales become tiny enough, there arise multiple
avenues for an irreversible conversion from magnetic to thermal energy.

Although there is no universal agreement about the best way to
quantitatively model the above chain of processes, we note that the
idea of a {\em turbulent cascade} has been quite successful in predicting
the heating rates and intermittency properties of the corona
\citep[e.g.,][]{vB86,Go00,Rp08,vB11,vB14,Db12,Ke15}.
In a plasma with a strong magnetic field, MHD turbulence is modulated
by Alfv\'{e}n waves, which propagate in both directions along the field
and interact with one another nonlinearly.
Observations show that both Alfv\'{e}n-like (incompressible)
and acoustic-like (compressible) waves appear to be present in the
corona \citep{Of99,KP12,Th13,Li15},
so it is worthwhile considering the effects of both on its heating.

Thus, in this paper we consider two sources of heat:
(1) Alfv\'{e}n waves that dissipate through a turbulent cascade, and
(2) compressive waves that dissipate via shocks and heat conduction.
The total volumetric heating rate is given by
\begin{equation}
  Q_{\rm heat} \, = \, Q_{\rm A} + Q_{\rm S}  \,\, ,
\end{equation}
where the subscripts ``A'' and ``S'' refer to Alfv\'{e}n and
compressive (sound) waves, respectively.
\citet{Nu13} suggested that the occurrence of down loops could be
caused by the conversion of Alfv\'{e}n waves into compressive modes
at the bases of loops with weak magnetic fields (i.e., values of the
plasma $\beta \gtrsim 1$).
We characterize this process with two free parameters: the total
wave energy density injected at the base of the loop ($U_{\rm tot}$)
and the fraction of initial Alfv\'{e}n waves that are converted into
compressive waves ($f$).
This gives the initial energy densities at the base of the loop as
\begin{equation}
  U_{\rm S,0} \, = \, f \, U_{\rm tot}
  \,\,\,\,\,\,\,\,\,\, \mbox{and} \,\,\,\,\,\,\,\,\,\,
  U_{\rm A,0} \, = \, (1-f) \, U_{\rm tot}  \,\,\, .
  \label{eq:convert}
\end{equation}
The remainder of this section describes how we compute the heating
rates $Q_{\rm A}$ and $Q_{\rm S}$ associated with each source of
wave energy.

\subsection{Incompressible Alfv\'{e}nic Turbulence}
\label{sec:heat:alf}

We begin with a model in which Alfv\'{e}n waves propagate up from
the photosphere and gradually ``feed'' a turbulent cascade.
This can only happen when there are counter-propagating wave packets
that can collide with one another \citep{Ir63,Kr65,HN13}.
This situation is clearly present in a coronal loop with both footpoints
being jostled by convective motions.
A long series of analytic models, computer simulations, laboratory
experiments, and comparisons with {\em in~situ} heliospheric plasma has
led us to generally believe that Alfv\'{e}n waves damp out due to
turbulence and produce heat at roughly the following rate,
\begin{equation}
  Q_{\rm A} \, \approx \, \rho \,\,
  \frac{Z_{+}^2 Z_{-} + Z_{-}^2 Z_{+}}{\lambda_{\perp}}
\end{equation}
\citep[see, e.g.,][]{Hs95,ZM90,Mt99,Dm02,Br08,Ch11,CvB12,vd14}.
The quantity $\lambda_{\perp}$ is a transverse correlation length,
or turbulent outer scale.
$Z_{\pm}$ are the Elsasser amplitudes that characterize the strength
of turbulent eddies propagating in both directions along the field line.
A symmetric closed loop will experience equal contributions of
Alfv\'{e}n waves from both directions, and we can assume
$Z_{+} = Z_{-}$.
In this case, the transverse root-mean-square velocity amplitude is
given by $v_{\perp} = Z_{\pm}/\sqrt{2}$, and the basal energy density
of Alfv\'{e}n waves can be estimated as
$U_{\rm A,0} = \rho_{0} v_{\perp 0}^{2}$.

We assume that the loss rate of waves is sufficiently weak such that
the rms velocity amplitude throughout the loop can be modeled with
wave flux conservation, which for hydrostatic loops is equivalent to
$v_{\perp} \propto \rho^{-1/4}$.
Additionally, we assume that the correlation length scales with the
radius of the flux tube \citep{Ho86} such that
$\lambda_\perp \propto A^{1/2} \propto B^{-1/2}$.
In terms of the free parameters $f$ and $U_{\rm tot}$, we can write
\begin{equation}
  Q_{\rm A} \, = \, \left[
  \frac{\alpha \, (1-f)^{3/2} U_{\rm tot}^{3/2}}
  {\rho_{0}^{1/2} \lambda_{\perp 0}} \right]
  \left( \frac{\rho}{\rho_0} \right)^{1/4}
  \left( \frac{B}{B_0} \right)^{1/2}
\end{equation}
where the subscript 0 refers to the value of each quantity
at the footpoints of the loop in the transition region, such that the
quantity in the square brackets above can be called the basal Alfv\'enic
heating rate $Q_{\rm A,0}$.
The constants $\alpha = 0.85$ and $\lambda_{\perp 0} = 1.1$~Mm
were adopted from the semi-empirical turbulence model of
\citet{CvB12}.
Note that specifying $Q_{\rm A}(s)$ requires knowing the solution
for the density $\rho(s)$, which is an output of the modeling process.

\subsection{Mode Conversion to Compressive Waves}
\label{sec:heat:comp}

\citet{Nu13} suggested that some fraction of the Alfv\'{e}n wave energy
may be converted, via one or more nonlinear mechanisms,
into compressive and longitudinal wave energy.
The resulting fluctuations are similar to acoustic waves (i.e.,
slow-mode magnetosonic waves in plasmas with $\beta < 1$), but are
not identical \cite[see, e.g.,][]{Ho71,Nk98,Bg02,Bg06,Kg07,CW15}.
Still, they behave in many ways like acoustic waves, in that they
exhibit oscillations in both density ($\delta \rho$) and
velocity parallel to the field ($\delta v_{\parallel}$).

We assume that the mode conversion occurs in the chromosphere and
lower TR, and at the lower boundary of our modeled loop there is a
compressive wave energy density $U_{\rm S,0}$ given by
Equation~(\ref{eq:convert}).
The compressive waves obey a conservation equation discussed in
more detail in Section 4.1 of \citet{CvB07},
\begin{equation}
  \frac{1}{A} \frac{\partial}{\partial s} \left( c_{s} A U_{\rm S} \right)
  \, = \, -Q_{\rm S} \, = \, -\frac{c_{s} U_{\rm S}}{L_{\rm damp}}
  \label{eq:dampS}
\end{equation}
where $c_s$ is the sound speed,
\begin{equation}
  c_{s} \, = \, \sqrt{\frac{\gamma P}{\rho}}
  \, = \, \sqrt{\frac{\gamma C_1}{C_2} \, \frac{k_{\rm B} T}{m_p}}
\end{equation}
where we assume $\gamma = 5/3$.
For now the damping length $L_{\rm damp}$ is assumed to be constant
and is treated as a free parameter.
In Section~\ref{sec:conc} we compare the most realistic values of
$L_{\rm damp}$ with predictions from the theory of shock steepening
and conductive dissipation.

By defining the quantity $Y \equiv c_{s} A U_{\rm S}$, we solve
Equation (\ref{eq:dampS}) as
\begin{equation}
  Y(s) \, = \, Y_{0} \, \exp \left( - \frac{s}{L_{\rm damp}} \right)
  \label{eq:Ys}
\end{equation}
between the base ($s=0$) and the apex of the loop ($s=L/2$).
This gives an explicit form for the heating rate due to compressive
waves,
\begin{equation}
  Q_{\rm S} \, = \, \frac{c_{s,0} \, f \, U_{\rm tot}}{L_{\rm damp}}
  \, \frac{A_0}{A(s)} \,
  \exp \left( - \frac{s}{L_{\rm damp}} \right)  \,\, .
  \label{eq:compheat}
\end{equation}
The heating rate $Q_{\rm S}(s)$ can be written explicitly without
knowing the solutions for density or temperature along the loop.
However, writing the energy density $U_{\rm S}$ requires knowing
$c_{s}(s)$, and thus also $T(s)$.
The velocity amplitude of the compressive waves is defined as
\begin{equation}
  v_{\parallel}(s) \, = \, \sqrt{\frac{3 \, U_{\rm S}}{\rho}}
\end{equation}
where the factor of 3 contains the assumption that compressive waves
rapidly steepen into a sawtooth-shaped train of shock waves
\citep[see][]{SS72,CvB07}.

\begin{figure}
\epsscale{1.18}
\plotone{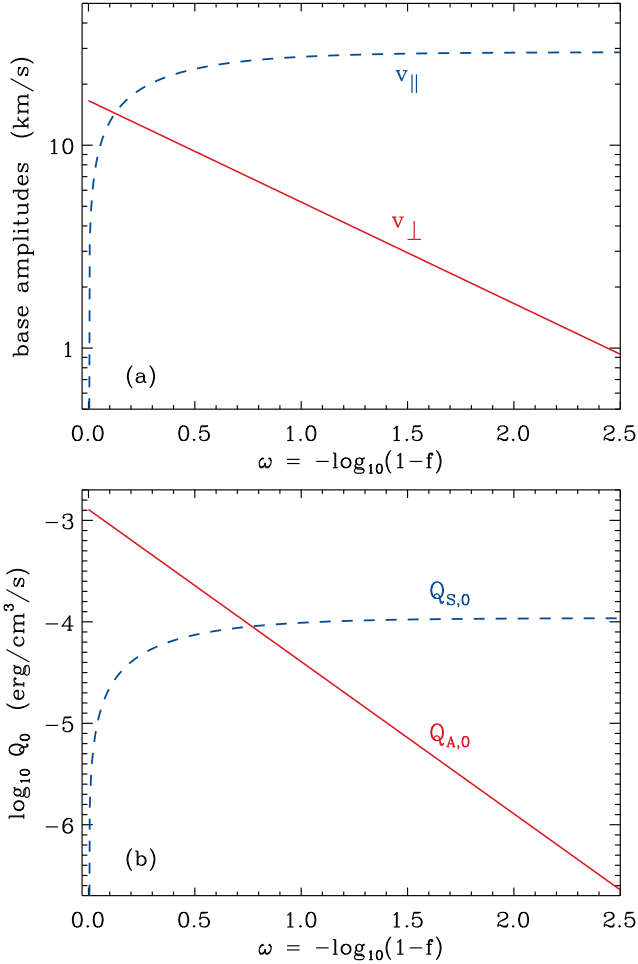}
\caption{(a) Basal velocity amplitudes and (b) basal heating rates
plotted versus mode conversion parameter $\omega$.
Alfv\'{e}nic (red solid curves) and compressive (blue dashed curves)
wave properties are compared against one another for a fixed total
wave energy density $U_{\rm tot} = 0.1$ erg cm$^{-3}$.
\label{fig04}}
\end{figure}

Figure~\ref{fig04} compares the basal amplitudes and heating rates
of Alfv\'{e}nic and compressive waves with one another for a range
of values for the mode conversion fraction $f$.
We find it useful to describe the conversion fraction with the
associated parameter $\omega \equiv -\log_{10} (1-f)$.
The parameter $\omega$ increases monotonically with $f$, with both
quantities starting at zero together.
However, as $f \rightarrow 1$, $\omega \rightarrow \infty$.
Integer values of $\omega$ give the number of 9's in the decimal
representation of $f$ (i.e., $\omega = 3$ corresponds to $f = 0.999$).
Thus, the region near $f \approx 1$ is stretched out and resolved
better with $\omega$.

Figure~\ref{fig04} shows how Alfv\'{e}n waves dominate at low values
of $\omega$ and compressive waves dominate at high values.
These models assumed $U_{\rm tot} = 0.1$ erg cm$^{-3}$,
$T_{0} = 10^{4}$~K, and $P_{0} = 0.05$ dyn cm$^{-2}$.
For these representative parameters, the velocities at the TR are
of order 20--30 km/s, which is representative of observational results
from nonthermal line broadening \citep[e.g.,][]{Bj11,HS14}.
Note that $U_{\rm A,0} = U_{\rm S,0}$ when $f=0.5$.
However, for the parameters assumed here, the heating rates
$Q_{\rm A,0}$ and $Q_{\rm S,0}$ do not become equal until $f=0.83$.
In other words, for low values of $f$ it is possible for
$Q_{\rm A,0} \gg Q_{\rm S,0}$ even when the velocity amplitudes are
comparable to one another.

\section{Numerical Methods}
\label{sec:method}

We developed a new set of computational tools to solve for the
time-steady properties of coronal loops.
The Python language was chosen for its well-tested algorithms
(e.g., numerical quadrature and the solution of ordinary differential
equations), and for its overall flexibility.
Below we describe the loop boundary conditions
(Section~\ref{sec:method:bc}), the adopted solution procedure
(Section~\ref{sec:method:solve}), and the stability of the solutions
(Section~\ref{sec:method:stab}).

\subsection{Boundary Conditions}
\label{sec:method:bc}

At the coronal base ($s=0$), we specify the boundary conditions
on the gas pressure ($P_0$) and temperature ($T_0$).
We fix $T_{0} = 10^{4}$~K and allow $P_0$ to vary as an
output of our method (see below).
In order to determine additional boundary conditions, we find it
convenient to introduce the quantity
\begin{equation}
  \Psi \, = \, A \, \kappa(T) \, \frac{dT}{ds}  \,\, .
  \label{eq:Psidef}
\end{equation}

At the apex of the loop ($s=L/2$), we take advantage of the assumed
symmetry as shown in Figure~\ref{fig01} and assume
\begin{equation}
  \left. \frac{dT}{ds} \right|_{s=L/2} \, = \,
  \left. \frac{d\Psi}{ds} \right|_{s=L/2} \, = \, 0 \,\, .
  \label{eq:upperbc}
\end{equation}

Determining the bottom boundary condition on the temperature gradient
$dT/ds$ (i.e., on the quantity $\Psi$) requires slightly more care
than just choosing a representative value.
This quantity depends sensitively on the modeled
thermal energy balance at $T_0$.
Following \citet{Hm82} and \citet{SvB05}, we assume that in the upper
chromosphere the local heating term $Q_{\rm heat}$ is relatively
unimportant in comparison to the conduction and radiative cooling
terms.
We also simplify the equations in the vicinity of the transition
region by using a generalized power-law form of
the radiative loss function, 
\begin{equation}
  \Lambda(T) \, = \, \Lambda_{0} T^{\gamma} \,\, ,
\end{equation}
for constant values of $\Lambda_0$ and $\gamma$.
We also use a simpler version of the conductive flux,
\begin{equation}
  F \, = \, -\kappa_{0} T^{5/2} \frac{dT}{dz}
\end{equation}
that includes only the electron conductivity.
The thermal balance equation therefore becomes
\begin{equation}
  \frac{dF}{dz} \, = \, -n_{e}^{2} \, \Lambda_{0} T^{\gamma}  \,\, .
  \label{eq:apthermeq}
\end{equation}
Multiplying both sides of Equation~(\ref{eq:apthermeq}) by $F$,
then dividing by $dT/dz$, gives a straightforwardly integrable form
\begin{equation}
  F \, dF \, = \, \psi \, T^{\gamma+(1/2)} \, dT
\end{equation}
where we assume that $\psi = \kappa_{0} \Lambda_{0} (n_{e}T)^2$,
proportional to the pressure squared, is a constant across the
chromosphere and TR.
Integrating both sides yields
\begin{equation}
  \frac{1}{2} \, \left( F_{2}^{2} - F_{1}^{2} \right) \, = \,
  \frac{2 \psi}{2 \gamma + 3} \left[ 
  T_{2}^{\gamma+(3/2)} - T_{1}^{\gamma+(3/2)} \right] \,\, ,
\end{equation}
which can be simplified further by considering that $T$ and $|F|$ are
both increasing rapidly with increasing height, such that
$T_2 \gg T_1$ and $|F_2| \gg |F_1|$.
This allows us to eliminate the ``2'' subscript and write
\begin{equation}
  F^{2} \, = \, \frac{4 \psi \, T^{\gamma + (3/2)}}{2\gamma+ 3}
  \,\, .
\end{equation}
Using the above definitions, this gives
\begin{equation}
  \frac{dT}{dz} \, = \, \sqrt{\left( \frac{4}{2\gamma + 3}
  \right) \, \frac{|Q_{\rm rad}|\, T}{\kappa}} \,\, ,
  \label{eq:dTbc}
\end{equation}
and we calculate $\gamma = 9.9$ from our tabulated values of the
radiative loss function at $T_{0} = 10^{4}$~K.
We use this value in Equations (\ref{eq:Psidef}) and (\ref{eq:dTbc})
to obtain $\Psi_0$ at the lower boundary.

\subsection{Solving for Time-Steady Coronal Properties}
\label{sec:method:solve}

We solve for the density and temperature throughout the loop using
a similar iterative method as \citet{SvB05}.
Thermal energy conservation is a second-order differential
equation, which we separated into two first-order coupled equations:
Equation (\ref{eq:Psidef}) and
\begin{equation}
  \frac{d\Psi}{ds} \, = \, -A \, \left( Q_{\rm rad} +
  Q_{\rm heat} \right)  \,\, .
  \label{eq:dPsi}
\end{equation}
These coupled equations were integrated from the base ($s=0$) to
the apex ($s = L/2$) using first-order Euler steps
to obtain $T(s)$ and $\Psi(s)$.
However, both $Q_{\rm rad}$ and $Q_{\rm heat}$ depend on the coronal
density as well as on temperature.
To obtain $\rho(s)$, we first integrated Equation~(\ref{eq:f})
for $P(s)$ and also used Equation~(\ref{eq:eos}).
Doing this requires an initial guess for the temperature.
This was given by an analytic solution to a balance between a
constant coronal heating rate $Q_{\rm heat}$ and a conduction rate
$Q_{\rm cond}$ dominated by the $\kappa_0$ electron term.
In Cartesian geometry ($A = \mbox{constant}$), this is given by
\begin{equation}
  T_{\rm guess} (s) \, = \, ( T_{\rm max}-T_{0} ) \,
  \left[ 1 - \left( \frac{z}{z_{\rm max}} - 1 \right)^{2}
  \right]^{2/7} + \, T_{0}  \,\, .
  \label{eq:inittemp}
\end{equation}
For this first guess, we assumed representative chromospheric and
coronal values $T_{0} = 10^{4}$~K and $T_{\rm max} = 10^{6}$~K.

After creating the above initial temperature function, we
also chose an initial guess for $P_{0}$ and proceeded to integrate
Equations (\ref{eq:Psidef})--(\ref{eq:dPsi}).
In general this initial solution does not satisfy the upper boundary
conditions at the loop apex (see above), so we used a Newton-Raphson
root finding algorithm to adjust $P_0$ toward a value at which
Equation (\ref{eq:upperbc}) is satisfied.
We used a fixed initial guess of $P_{0} = 0.01$~dyn cm$^{-2}$ for
the Newton-Raphson method.
This value is noticeably lower than the typical base pressure found
for the loops, but we find that values of the base pressure that are
too high tend to cause divergent solutions.
In rare cases when the Newton-Raphson method failed to converge to a
solution, we found the root with a more stable but significantly slower
bisection method (searching between bounding values of
$P_{0} = 0.001$ and 10 dyn cm$^{-2}$).

As the numerical iteration progresses, the tabulated solutions
for $T(s)$ and $P(s)$ are updated and the most recent values are used
as guesses in successive rounds of integrating Equations (\ref{eq:f})
and (\ref{eq:Psidef})--(\ref{eq:dPsi}).
The finite-difference tables contain 1000 values of $s$
between the base and apex, and they use a nonuniform grid determined
by the function
\begin{equation}
  s \, = \, \frac{L}{2} \, \left( \frac{n}{999} \right)^{4}
\end{equation}
for $0 \leq n \leq 999$.
Grid zones are made intentionally dense near $s = 0$ in order to capture
the rapid temperature changes seen near the base of the loop.
We tested the adequacy of these discrete grid parameters (and the
use of first-order Euler integration) everywhere in the interior
of the loop grid by considering the sum of the
three heating/cooling terms in Equation (\ref{eq:mainthermal})
divided by the absolute value of the largest of $Q_{\rm heat}$, 
$Q_{\rm rad}$, or $Q_{\rm cond}$.
When averaged over the length of the loop, this testing ratio was
found to almost never exceed 1\% for both short and long loops, with
the average usually closer to 0.6-0.9\% for both cases.
At worst, for some combinations of parameters this ratio would reach 15\% 
at some point along the short loops and 27\% at some point along the long 
loops. 
The instances with the relatively high test ratio tended to coincide
with low-$\omega$ and high-$U_\textrm{tot}$ loops that would have
unrealistically high peak temperatures and temperature gradients
near the top of a quiescent corona.
The usual low values of the test ratio, especially for loops
matching the observational data,
indicates a comparable level of fractional accuracy in the other
parameters.

Because of the possibility of large relative changes from one
iteration to the next, we used an undercorrection scheme for the
temperature.
This technique was adopted from the ZEPHYR code \citep{CvB07}, and
the value at each distance $s$ is updated using
\begin{equation}
  T(s) \, = \, T_{\rm old}(s)^{N} \, T_{\rm new}(s)^{1-N}
\end{equation}
where $T_{\rm old}$ is the previous iteration's solution and
$T_{\rm new}$ is the direct output of the numerical integration
for the current iteration.
The constant $N$ is a parameter between 0 and 1 that we set to 0.65 for
most cases.
For loops that tend to diverge during early iterations, $N$ is
first set to 0.2 and then gradually raised to 0.5.
Once $T(s)$ is updated, the table of $P(s)$ values is updated, and
the process is repeated
until the relative temperature difference between one iteration and
the next is less than a thousandth of a percent for ten sample grid
points spaced evenly through the grid, from $n\, =\, 99$ to $n\, =\, 999.$

\subsection{Stability Considerations}
\label{sec:method:stab}

In order to determine whether the time-steady solutions are stable
to small perturbations, we test for unstable gravitational
stratification.
According to \citet{As01}, a hydrostatic loop becomes
{\em dynamically unstable} if
\begin{equation}
  \frac{\partial \rho}{\partial s} > 0
  \label{eq:gravstrat}
\end{equation}
in the region $0 \leq s \leq (L/2)$.
Often, it is not necessary to apply this criterion because unstable
solutions tend to be unphysical in other ways.
In almost all cases, our models exhibit monotonically decreasing
density with increasing $s$.
If the coronal heating is concentrated too strongly at the base,
Equation (\ref{eq:gravstrat}) may end up being satisfied, but these
models tend to have insufficient heating toward the apex to maintain
a time-steady hot corona.
The iteration process in this case leads to negative temperatures,
clearly indicating an unphysical solution.

There have been a number of different {\em thermal instabilities}
proposed that could disrupt some coronal loops and give rise to
prominence condensation or ``coronal rain''
\citep[e.g.,][]{Pr78,HR79,Go90,CR91,Mk08,Ss12,Ss15,An15}.
This process---in which the loop is stretched or twisted beyond a
stable length and cools catastrophically to chromospheric
temperatures---is just one of several proposed to explain the
highly dynamic and intermittent ultraviolet and X-ray emission seen
from loops.
However, we defer the study of these kinds of instabilities to future work,
since the large-scale quiescent loops modeled in this paper appear to
have parameters quite distinct from the (mainly low-lying) unstable
loops in the studies cited above.

\section{Results}
\label{sec:results}

\subsection{Model Grids}
\label{sec:results:grids}

We consider loops of two different lengths, both given in Table~1,
and we refer to them as ``long'' and ``short'' loops with reference
to the categories defined by \citet{Nu13}.
The short loops are characterized by $\theta = 8\arcdeg$ and
$L = 0.333 \, R_{\odot}$, and
the long loops have $\theta = 25\arcdeg$ and
$L = 1.584 \, R_{\odot}$.
To span the full parameter space of possible loop solutions for each
length, we fill a three-dimensional grid of values for $f$,
$U_{\rm tot}$, and $L_{\rm damp}$.
Twenty different values for $f$ are used, with ten values evenly
spaced between 0 and 0.9 (i.e., $0 \leq \omega \leq 1$) and another
ten values evenly spaced in $\omega$ between 1.3 and 4.0.
Ten different values are used for $U_{\rm tot}$ that are evenly spaced
on a logarithmic scale bounded by 0.01 and 3.16 erg cm$^{-3}$.
For most values of $f$ we use 13 different values of $L_{\rm damp}$.
However, when $f = 0$, the compressive heating term described in
Equation (\ref{eq:compheat}) vanishes and the coronal heating
term is purely Alfv\'{e}nic.
It is therefore only necessary to simulate the loops once for each value
of $U_{\rm tot}$ when $f = 0$ since the damping length no longer
affects the heating rate.
For all other values of $f$, the 13 values of $L_{\rm damp}$ are
evenly spaced between 0.004 and 0.04 $R_{\odot}$.
In total, 4960 loops are simulated, half of which are short and
half of which are long.

In order to draw as close a parallel as possible to the findings of
\citet{Nu13}, we often average values along the loop ``legs''
at the heights probed by the DEMT technique.
The legs are defined to span heights between 0.03 and
0.2 $R_{\odot}$, with the larger value replaced by $z_{\rm max}$
for the short loops that do not extend as high as 0.2 $R_{\odot}$.
Specifically, we compute mean temperatures $\langle T \rangle$
in the legs, and we use the slope of a linear fit to $T(z)$ in the
legs to determine a mean temperature derivative
$\langle dT/dz \rangle$.
Models with a positive slope are characterized as ``up loops'' and
those with a negative slope are ``down loops.''
We note, however, that this one parameter does not always accurately
characterize the full spatial dependence of $T(s)$.

\subsection{Representative Solutions}
\label{sec:results:rep}

\begin{figure}
\epsscale{1.15}
\plotone{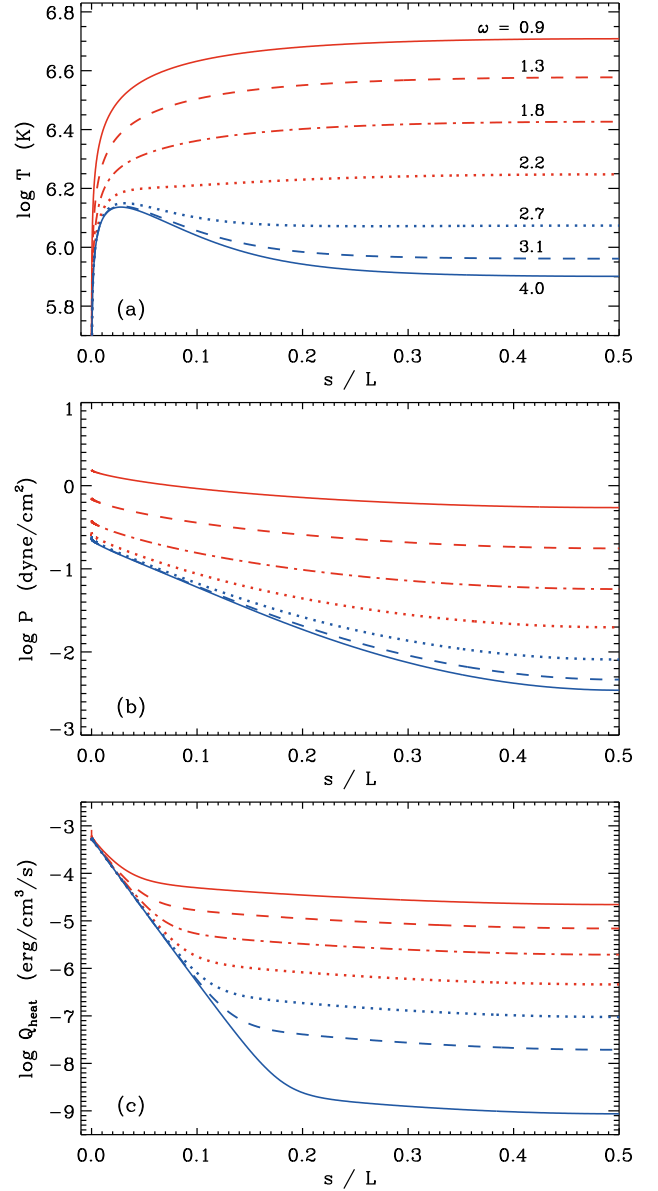}
\caption{Spatial dependence of (a) temperature, (b) gas pressure,
and (c) total volumetric heating rate, as a function of distance $s$
traced along long loops.
Fixed values of $U_{\rm tot} = 1$ erg cm$^{-3}$ and
$L_{\rm damp} = 0.025 \, R_{\odot}$ were used with a range of
choices for $\omega$ (i.e., mode conversion fraction $f$); see curve
labels in panel (a).
Up and down loops are plotted in red and blue, respectively.
\label{fig05}}
\end{figure}

Figure \ref{fig05} shows a selection of temperature and pressure profiles
for long loops ($L = 1.583 \, R_{\odot}$) with a range of $\omega$ values
and fixed values of $U_{\rm tot} = 1$ erg cm$^{-3}$ and
$L_{\rm damp} = 0.025 \, R_{\odot}$.
Increasing the amount of mode conversion (i.e., larger values of $\omega$)
decreases the total amount of heat deposited along the loop, but it also
makes the heating rate more strongly peaked at the footpoints.
\citet{AS02} showed that a sufficiently steep decrease in $Q(s)$ indeed
gives rise to a ``down loop'' with a local minimum in temperature at
the apex.

We determined that stable down loops become possible when the mode
conversion from Alfv\'{e}nic to compressive waves is strong
(i.e., $\omega \gtrsim 2$).
Down loops tend to be seen only for a finite range of $U_{\rm tot}$
values, and the bounds of this range vary inversely with $L_{\rm damp}$.
For values of $U_{\rm tot}$ smaller than this finite range, loops
remain ``normal'' (i.e., with monotonically increasing temperatures)
for all values of $\omega$.
For values of $U_{\rm tot}$ larger than this finite range, the amount
of deposited energy at the base appears to be too large for stable
coronal loops to exist, and the code is unable to find solutions with
$T > 0$ everywhere.\footnote{%
Of the 4960 models, the code was unable to find stable
solutions for 571 (11.5\%) of the parameter combinations.}
Detailed maps of the parameter space are shown below.

A number of loops are poorly described as ``up'' or ``down''
when their slope $\langle dT/dz \rangle$ is only taken from a linear
fit along the legs as described above.
In some cases, the peak temperature occurs below the apex, but it is
too high for it to be registered as a down loop.
Of the 1940 short loops recorded as ``up'' loops, 214 (11\%) have
maximum temperatures before the apex, while the same is true for
189 (8.9\%) of the 2115 long ``up'' loops. 
It is possible that these kinds of loops are not reported by \citet{Nu13}
because they only considered loops for which it was possible to create
a high-quality linear fit (i.e., with $R^2 > 0.5$) from the DEMT data.

\begin{figure}
\epsscale{1.18}
\plotone{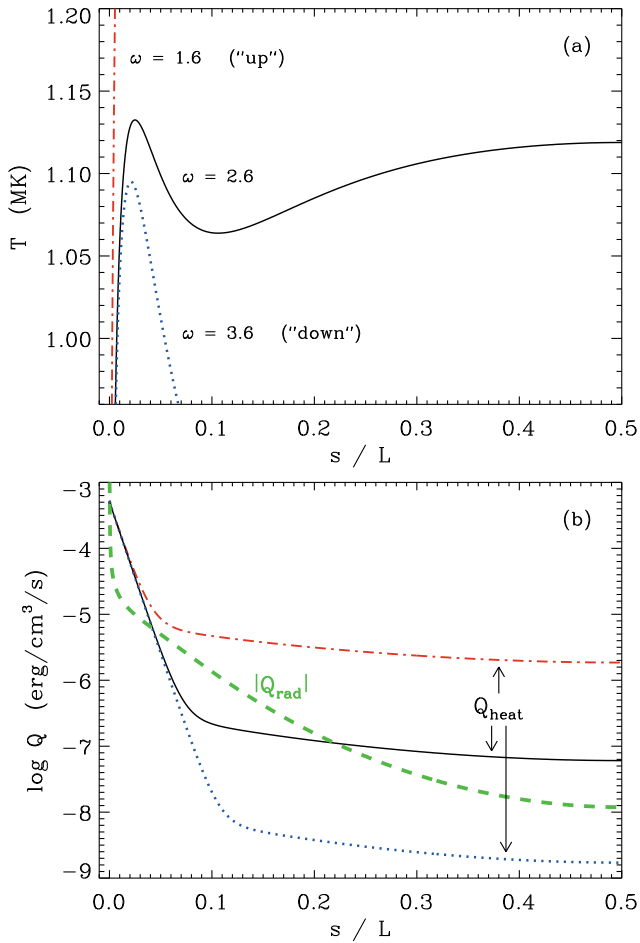}
\caption{Spatial dependence of (a) temperature, and (b) heating/cooling
terms in Equation (\ref{eq:mainthermal}), for a set of long loop models
with $U_{\rm tot} = 0.6$ erg cm$^{-3}$,
$L_{\rm damp} = 0.016 \, R_{\odot}$, and a range of values for
$\omega$ (see curve labels in panel a).
Red, black, and blue curves in panel (b) show $Q_{\rm heat}$, and
the green dashed curve shows $|Q_{\rm rad}|$ for only the middle
case of $\omega = 2.6$.
\label{fig06}}
\end{figure}

In Figure~\ref{fig06} we also illustrate a transitional nonmonotonic
temperature profile that occurs for some long-loop models.
Standard up and down loops exhibit one and two temperature maxima
(when measured from footpoint to footpoint), respectively.
However, these ``hybrid'' loops exhibit {\em three} temperature peaks;
i.e., they have a local maximum at the apex, local minima below the apex,
and additional local maxima below them on either side.
Figure~\ref{fig06}(b) illustrates how these hybrid loop cases can occur.
Equation (\ref{eq:dPsi}) indicates that if $|Q_{\rm rad}| > Q_{\rm heat}$,
the second derivative of $T$ is positive and the temperature curve
near the peak is concave up, consistent with down loops.
If $|Q_{\rm rad}| < Q_{\rm heat}$, the curve is concave down
at the apex, consistent with up loops.
However, in Figure~\ref{fig06}(b) we see that for the hybrid loop case,
the ratio $|Q_{\rm rad}|/Q_{\rm heat}$ oscillates above and below 1
a number of times between the base and the apex.
Thus, despite the strong conduction that tends to smear out small-scale
thermal fluctuations along the field, these rare cases do demand multiple
minima and maxima in the time-steady $T(s)$.

\subsection{Statistical Trends}
\label{sec:results:trends}

\begin{figure*}
\epsscale{1.15}
\plotone{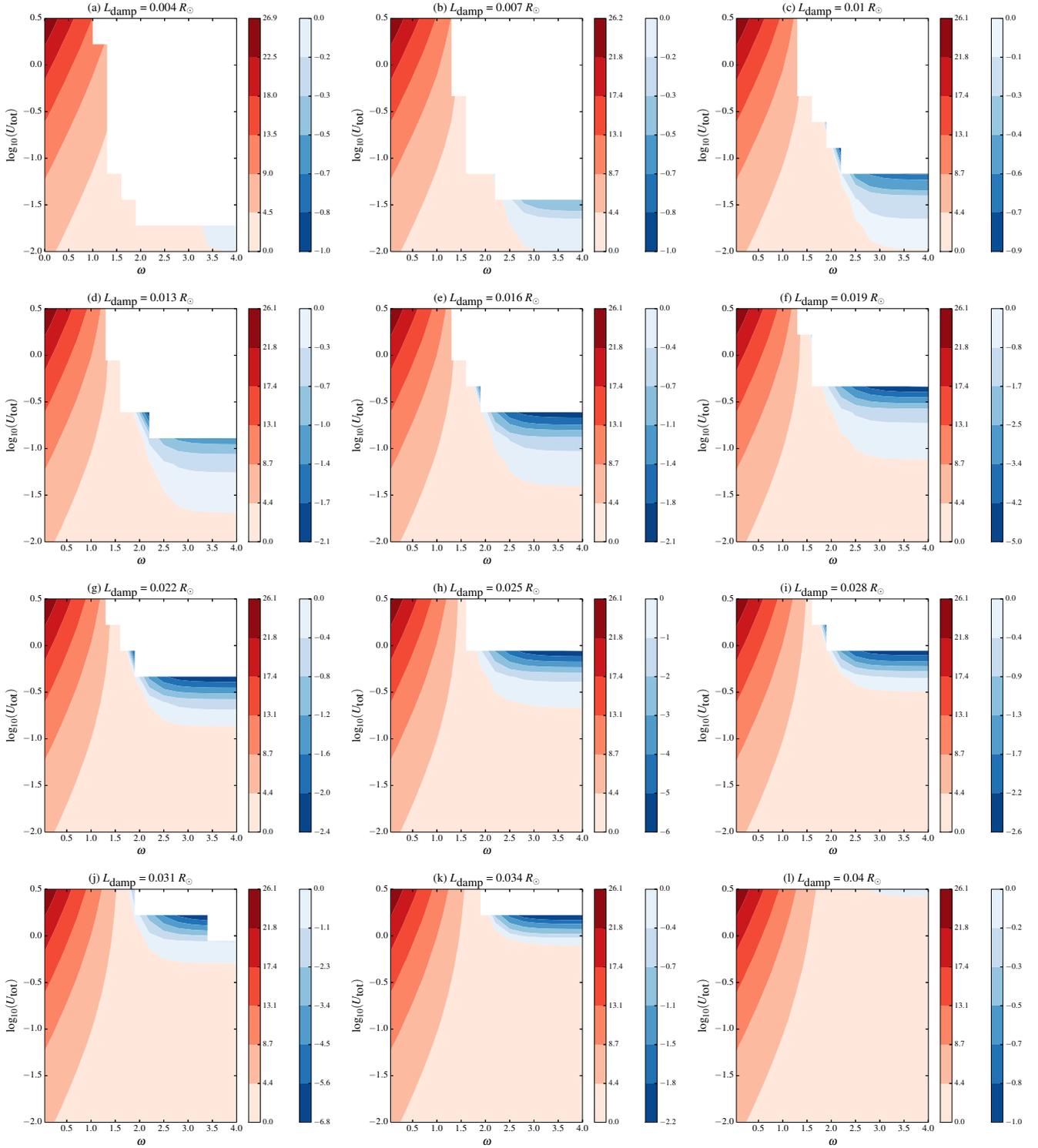}
\caption{Contour plots of $\langle dT/dz \rangle$ versus $\omega$
and $U_{\rm tot}$, for short-loop models with $L = 0.335 \, R_{\odot}$.
Each panel shows results for a fixed value of $L_{\rm damp}$.
Color scales for positive (red) and negative (blue) values of
$\langle dT/dz \rangle$ are unique for each panel, and the numbers
in color bars are given in units of MK~$R_{\odot}^{-1}$.
White regions denote parameters with no stable solutions.
\label{fig07}}
\end{figure*}

\begin{figure*}
\epsscale{1.15}
\plotone{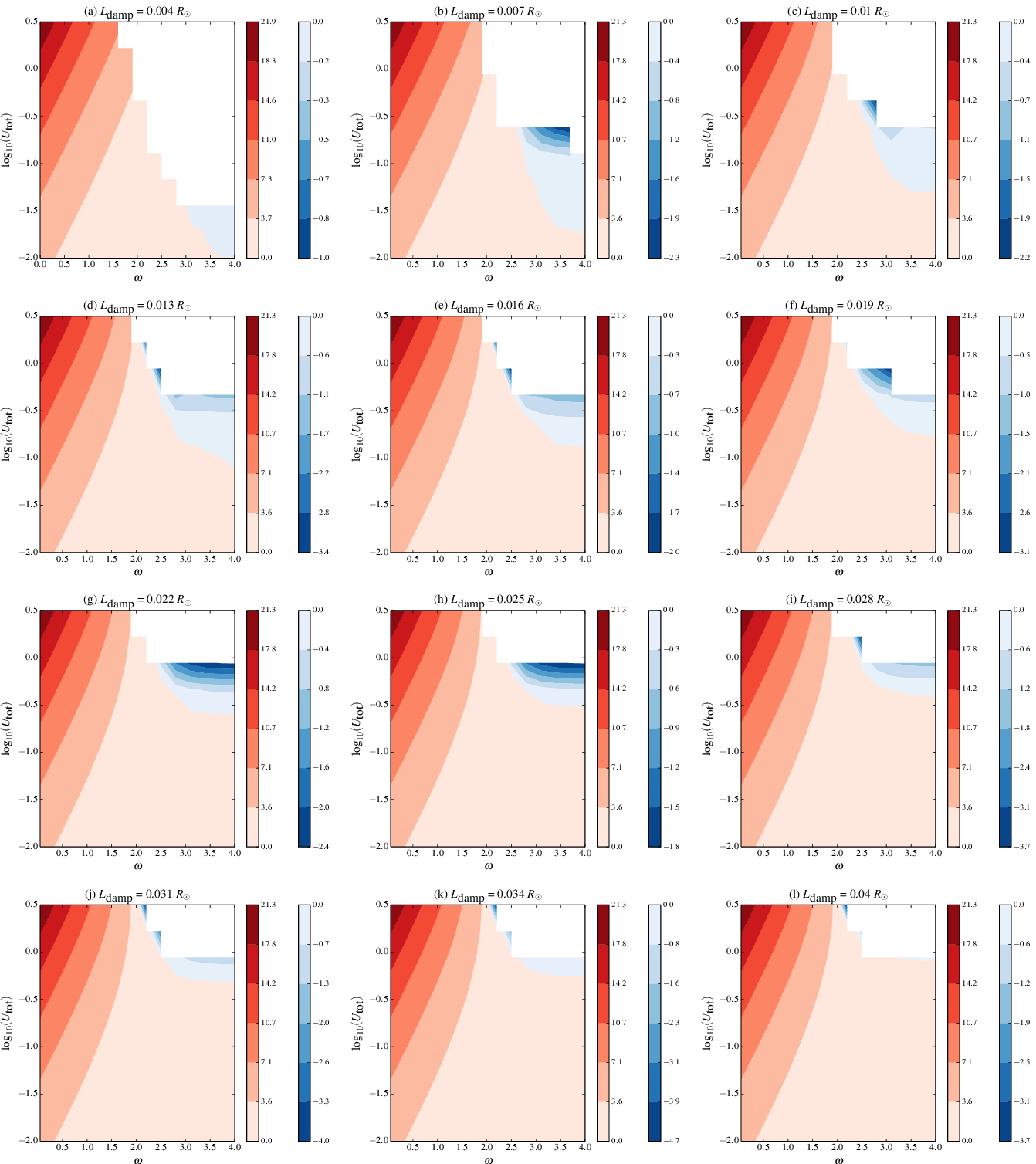}
\caption{Same as Figure~\ref{fig07}, but for the long-loop models
with $L = 1.583 \, R_{\odot}$.
\label{fig08}}
\end{figure*}

Figures \ref{fig07} and \ref{fig08} are contour plots that indicate
the mean temperature gradient $\langle dT/dz \rangle$ in units of
MK~$R_{\odot}^{-1}$.
Each panel plots $\langle dT/dz \rangle$ versus $\omega$ and $U_{\rm tot}$
for a fixed value of $L_{\rm damp}$.
Figure~\ref{fig07} shows results for the short loop models, and
Figure~\ref{fig08} shows results for the long loop models.
We continue the color scheme from Figures~\ref{fig05} and \ref{fig06}
\citep[and from][]{Nu13} in which up-loop parameters are in red and
down-loop parameters are in blue.

White areas in Figures \ref{fig07} and \ref{fig08}
indicate combinations of parameters for which the code did
not find physically realistic solutions.
This tends to occur for the largest values of both $\omega$ and
$U_{\rm tot}$; i.e., very strong basal heating and a rapid drop-off of
$Q_{\rm heat}$ with increasing height.
All cases of ``good'' solutions (not in white) were found to be also
gravitationally stable ($\partial \rho / \partial s < 0$) everywhere
from footpoint to apex.
This result disagrees somewhat with \citet{AS02}, who found some cases
of gravitationally unstable loops with positive temperatures at all
values of $s$.
These cases occurred for similar (i.e., rapidly decreasing) heating
functions as our no-solution cases, but it should be noted that
the exponential form for $Q_{\rm heat}(s)$ used by \citet{AS02}
was different from what is used here.

\begin{figure}
\epsscale{1.15}
\plotone{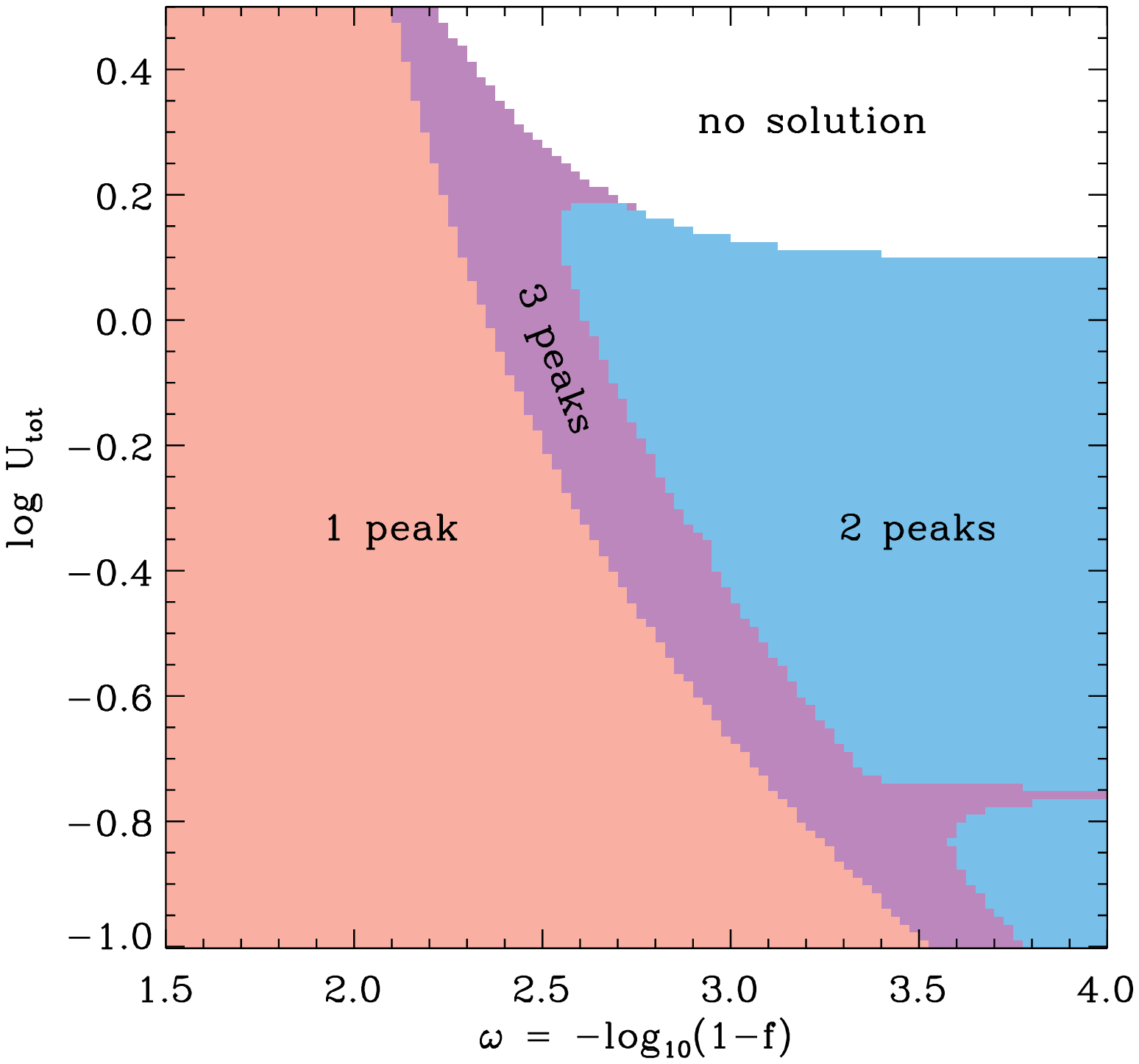}
\caption{Contour plots of the number of maxima in $T(s)$ measured from
footpoint to footpoint: with regions of one (red), two (blue), and
three (purple) shown alongside the region of no physical solutions
(white).  Models were computed for the long-loop case, with
$L_{\rm damp} = 0.034 \, R_{\odot}$.
\label{fig09}}
\end{figure}

As noted above, the sign of the observationally motivated mean gradient
$\langle dT/dz \rangle$ does not always convey an accurate picture of
the number of temperature minima and maxima.
Specifically, it does not reveal the region of parameter space that
produces ``hybrid loops'' (i.e., ones with three maxima in $T(s)$
from footpoint to footpoint) at all.
Thus, in Figure~\ref{fig09} we give an example of the same parameter
space shown in Figures \ref{fig07}--\ref{fig08}, but with a finer grid
and a color scale that denotes the exact topological nature of each
model's temperature profile.
This grid contained 12,000 models
that were only simulated until the
peak temperature had sufficiently converged:
100 values of $\omega$ and 120
values of $U_{\rm tot}$ that span the limits shown in Figure~\ref{fig09}.
The hybrid loops (3 peaks) occupy a narrow strip of parameter space
between the standard regions of up loops (1 peak) and down loops (2 peaks),
with some small-scale boundary structure that depends on the relative
magnitudes of the heating terms $Q_{\rm rad}$ and $Q_{\rm heat}$ as
discussed above.
The purple ``spurs'' that extend to the right in Figure~\ref{fig09} 
correspond to loops with extremely shallow minima beneath their apex.
In those cases the temperature variation at the top of the loop is almost
flat, and a plot of $T(s)$ would look nearly indistinguishable from a
two-peak down~loop.

\begin{figure}
\epsscale{1.18}
\plotone{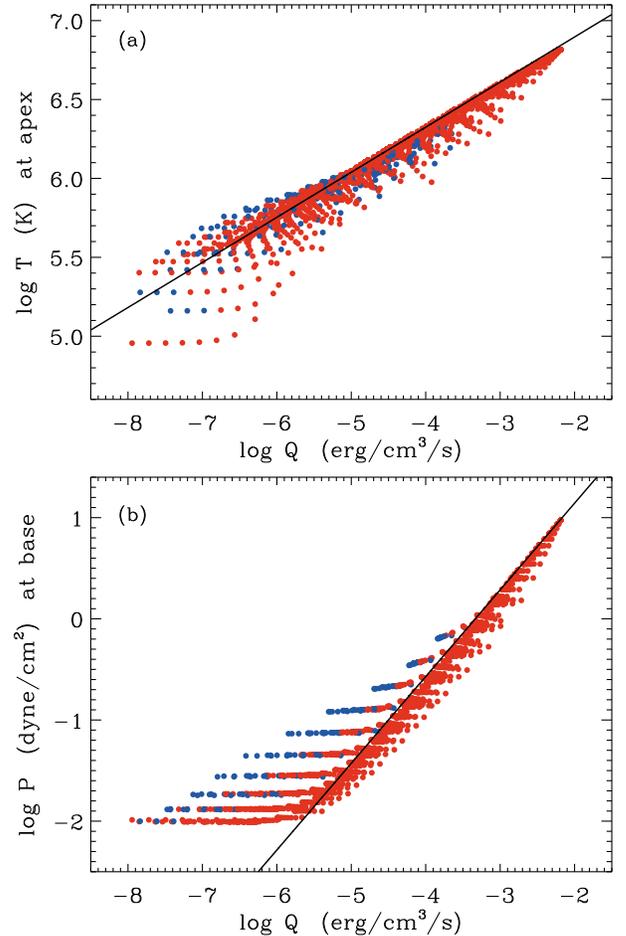}
\caption{Numerically computed apex temperature $T_{\rm max}$ and
base pressure $P_0$ for short up~loops (red points) and short down~loops
(blue points), plotted against mean heating rate $Q$ (see text).
RTV power-law predictions are shown with solid black lines.
\label{fig10}}
\end{figure}

The numerical results for apex temperature $T_{\rm max}$ and base
pressure $P_0$ can be compared straightforwardly to several existing
analytical models \citep[e.g.,][]{RTV,Se81,AS02,Ma10}.
For simplicity's sake we compare only with \citet[][hereafter RTV]{RTV},
who assumed a constant heating rate $Q$ versus distance along a loop with
constant pressure and cross section.
For standard choices of the radiative and conductive constants,
the RTV laws can be written as
\begin{displaymath}
  T_{\rm max} \, = \, 
1.302 \,\, \mbox{MK} \, 
  \left( \frac{Q}{\mbox{10$^{-4}$ erg cm$^{-3}$ s$^{-1}$}} \right)^{2/7} 
\end{displaymath}
\begin{equation}
  \times\left( \frac{L}{\mbox{100 Mm}} \right)^{4/7}
  \label{eq:rtvT}
\end{equation}
\begin{displaymath}
  P_{0} \, = \, 0.1467 \,\, \mbox{dyn cm$^{-2}$} \,\\
  \left( \frac{Q}{\mbox{10$^{-4}$ erg cm$^{-3}$ s$^{-1}$}} \right)^{6/7} 
\end{displaymath}
\begin{equation}
  \times\left( \frac{L}{\mbox{100 Mm}} \right)^{5/7} \,\, .
  \label{eq:rtvP}
\end{equation}
Because the heating rates used in this paper are monotonically decreasing
from footpoint to apex, it is not clear what value(s) of $Q$ to use in
Equations (\ref{eq:rtvT})--(\ref{eq:rtvP}).
After some experimentation, we found that using the geometric mean of the
basal and apex heating rates (i.e., the square root of their product)
produces reasonable agreement with the RTV predictions.
Figure~\ref{fig10} shows a model-by-model comparison for the
grid of short loops.
A comparable plot for the long loops looks similar to this one, but with
all values of $T_{\rm max}$ and $P_0$ shifted slightly upwards.
Note that the up~loops (red points) tend to be more RTV-like than
the down~loops (blue), since the latter often have a more pronounced
and complicated $s$-dependence to their heating rates and pressures.

\begin{figure}
\epsscale{1.18}
\plotone{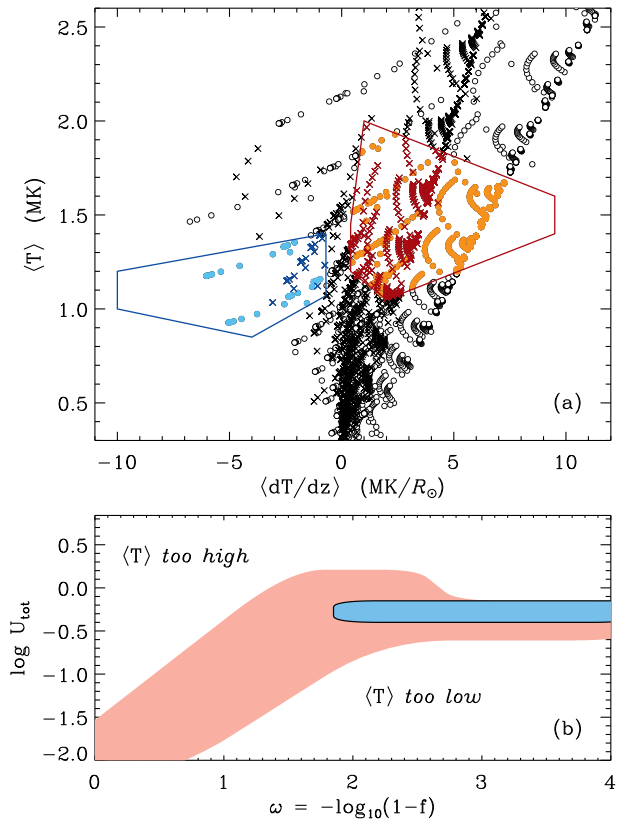}
\caption{(a) Scatter plot of fitted mean loop temperatures
$\langle T \rangle$ versus the mean temperature gradient
$\langle dT/dz \rangle$.
Up-loop solutions are shown for short loops (orange circles) and
long loops (dark red X's).
Down-loop solutions are shown for short loops (cyan circles) and
long loops (dark blue X's).
Outside the observed region of parameter space \citep{Nu13},
symbols are black.
(b) Regions of model parameter space (as in Figures
\ref{fig07}--\ref{fig09}) that correspond approximately to
the models inside the observed regions in panel (a).
Up-loop regions are in red and down-loop regions are in blue.
\label{fig11}}
\end{figure}

For a more direct comparison with the observational data,
Figure~\ref{fig11}(a) shows how $\langle T \rangle$ and
$\langle dT/dz \rangle$ compare against one another.
Polygonal outlines \citep[adapted from Figure 10 of][]{Nu13}
give an approximate indication of the observed region of parameter
space for up and down loops.
Out of the 4960 cases simulated in the short and long loop grids,
only 698 of them (14\%) fall inside the polygonal outlines.
Figure~\ref{fig11}(b) outlines the regions of parameter space
(in $U_{\rm tot}$ and $\omega$) corresponding to the these models.
There seems to be an intermediate range of values for $U_{\rm tot}$
(which changes slightly based on the mode coupling efficiency) required
to match the observations.
More wave power would result in too much heating (i.e.,
$\langle T \rangle > 2$~MK), and less wave power would give too little
($\langle T \rangle < 0.8$~MK).
There are no clear limiting values of $L_{\rm damp}$ that identify the
subset of models that agree with the observations.

Figure~\ref{fig11} gives the general impression that the measured
quiescent up~loops are produced when there is a {\em small} amount of
mode energy transfer from Alfv\'{e}n to compressive waves at the TR,
and the down~loops are produced when there is {\em large} mode
energy transfer.
For low values of $\omega$, the band of observationally appropriate
parameters for up~loops corresponds to basal values of the Alfv\'{e}n
velocity amplitude around 3--6 km~s$^{-1}$.
These values are quite a bit smaller than the canonical 20--30 km~s$^{-1}$
nonthermal amplitudes often seen in the brighter and more compact
active-region loops.
However, the large quiescent loops traced out by the DEMT technique
\citep{Hu12,Nu13} are believed to comprise a {\em diffuse coronal
background} that must be heated more weakly than the small and bright
loops that are easier to resolve in extreme ultraviolet images.

\citet{Nu13} measured the plasma $\beta$ ratio for the observed loops, where
\begin{equation}
  \beta \, = \, \frac{P}{B^2/8\pi} \,\, .
\end{equation}
The leg-averaged value $\langle \beta \rangle$ was found to be
systematically larger for down~loops (median values of 1--3) than for
up~loops (median values of 0.5--0.9).
\citet{Nu13} attributed this difference mainly to the fact that
down~loops tend to have weaker magnetic fields than up~loops.
They also speculated that the larger values of $\langle \beta \rangle$
in the down~loops should make them more susceptible to mode conversion
between Alfv\'{e}n and compressive waves.

Unfortunately, we cannot make a direct comparison with the observed
trends in $\beta$ because our modeled loops do not depend on an
absolute normalization for $B$.
However, we have estimated the value of $\beta_0$ at the TR
under the assumption that all loops share a common value of $B = 2$~G
at the TR.
In fact, we believe $\beta_0$ may be a more important parameter for
determining the degree of basal mode conversion than
$\langle \beta \rangle$ measured higher up in the corona.
With the above assumption about $B$,
the up~loops within the red polygonal outline in
Figure~\ref{fig11}(a) have a median value of $\beta_{0} = 0.636$,
and the corresponding down~loops have a median value of
$\beta_{0} = 1.32$.
This difference is attributable solely to differences in $P_0$
in the two populations of models, but it does go in the same direction
as the observed variation in $\langle \beta \rangle$.  
It should be noted that a similar trend in base pressures is also evident
in the measured mean values given in Table~1 of \citet{Nu13}.
Thus, it is unclear whether the observed differences in
$\langle \beta \rangle$ between the up and down loops can be attributed
primarily to differences in gas pressure or magnetic pressure.

\section{Discussion and Conclusions}
\label{sec:conc}

One key goal of this paper was to test the conjecture made by
\citet{Nu13} that mode conversion at the TR---from Alfv\'{e}nic to
compressive waves---can be responsible for the existence of coronal
loops with an ``inverted'' temperature structure.
We constructed a large grid of models with two sources of time-steady
coronal heating: a turbulent cascade of incompressible Alfv\'{e}n waves
(which generally produces heating that varies slowly with height)
and the rapid basal dissipation of compressive waves (whose heating
is highly peaked at the loop footpoints).
The production of stable loops with radially decreasing temperatures 
requires nearly all (i.e., $>$99\%) of the Alfv\'{e}n wave energy to
be converted to compressive waves that deposit their heat at the base.
This was anticipated in earlier parameter studies \citep{Se81,AS02},
but we have quantified how much wave-mode energy transfer is needed to
produce such highly peaked heating functions.

One aspect of the models that was relegated to free-parameter status
was the damping length $L_{\rm damp}$ of the compressive waves.
Our grid of models assumed values of $L_{\rm damp}$ that vary over an
order of magnitude between 0.004 and 0.04 $R_{\odot}$.
It is worthwhile to compare these values with predictions from the
theory of magnetoacoustic wave dissipation.
\citet{CvB07} modeled this by assuming two sources of energy loss:
(linear) thermal conductivity and (nonlinear) shock entropy evolution.
The latter was found to be more important in the chromosphere, and the
former was found to be more important in the corona.
Thus, we examine heat conductivity, which has an effective damping
length $\tilde{L}_{\rm damp} = c_{s}/2\gamma_{c}$, where $\gamma_c$ is
the linear damping rate due to electron heat conduction.
Making use of classical expressions for this transport coefficient
\citep[see also][]{HB73,Wh97},
\begin{displaymath}
  \tilde{L}_{\rm damp} \, \approx \, 0.0133 \, R_{\odot} \,
  \left( \frac{\rho}{10^{-16} \, \mbox{g cm$^{-3}$}} \right)
\end{displaymath}
\begin{equation}\times
  \left( \frac{T}{10^{5} \, \mbox{K}} \right)^{-1}
  \left( \frac{\Pi}{\mbox{5 min}} \right)^{2}
  \label{eq:Ldtheory}
\end{equation}
where $\Pi$ is the acoustic wave period.
When evaluated at the chromospheric base, this damping length is
very large (10--100 $R_{\odot}$), but it drops rapidly with
increasing height.
For typical coronal-loop apex properties, this theoretical
$\tilde{L}_{\rm damp}$ is of order $10^{-3} \, R_{\odot}$.
The height at which the value of $\tilde{L}_{\rm damp}$ matters
most to the models (Equations (\ref{eq:Ys})--(\ref{eq:compheat}))
is essentially $s \approx \tilde{L}_{\rm damp}$.
This is the distance above the TR at which the damping finally
gets a chance to reduce the wave energy density by a significant
amount.
Thus, we can identify a unique damping length by finding the location
at which $\tilde{L}_{\rm damp}$ (a decreasing function of distance
along the loop) and $s$ (an increasing function, by definition) become
equal to one another.

For each of the 798 models inside the polygons in Figure \ref{fig11},
we determined a unique conductive damping length $\tilde{L}_{\rm damp}$
using the above procedure.
We assumed a fiducial chromospheric value of $\Pi = 3$~minutes
\citep{Ny67,Ru03,Ju06}.
This produced a range of values that matched the initial assumptions
from our grid quite well.
The minimum, median, and maximum values of $\tilde{L}_{\rm damp}$
were 0.0052, 0.0072, 0.047 $R_{\odot}$, respectively.
If we had used larger periods, such as the 20--30 minute
values observed for off-limb intensity oscillations
\citep[e.g.,][]{Of99,Li15}, then Equation (\ref{eq:Ldtheory}) would
have given much larger damping lengths.
However, such long period waves may be susceptible to conversion to
other types of conductivity-dominated modes.
\citet{Bg06} showed that such modes can dissipate rapidly at the coronal
base, thus effectively reducing $\tilde{L}_{\rm damp}$ despite the
large values of $\Pi$.

Although the models presented in this paper have helped to refine the
connections between wave mode conversion and loop temperature structure,
there are many other improvements that would allow us to build a more
comprehensive understanding of heating in the closed corona.
For example, instead of assuming the compressive waves are generated
impulsively at the lower boundary, the models should include a
self-consistent description of the height-dependence of plasma parameters
in the chromosphere and TR, as well as the $\beta$-dependent physics of
mode conversion \citep[see also][]{CW15}.
Time-dependent simulations are beginning to show a broad diversity of
mode coupling processes in this complex environment \citep{MS14,Ar16}.
Lastly, our adopted rate of heating due to MHD turbulent cascade was
a relatively simple phenomenological scaling law, but there are others
\citep[e.g.,][]{Rp08,vB11,Ag13,Bd16} that may be more realistic.

\citet{Hu12} and \citet{Nu13} found that inverted-temperature loops
tend to occur most frequently at low heliographic latitudes, and that
they preferentially appear in the ``deep'' solar minimum.
Thus, the footpoint locations of down~loops correspond with some of the
weakest large-scale magnetic fields seen over the solar cycle.
It is not surprising that they may have higher plasma $\beta$ ratios
and greater amounts of wave mode conversion than the rest of the corona.
Trans-equatorial loops like these connect to the cusp regions of large
helmet streamers, which also exhibit $\beta \gtrsim 1$ due to their
weak fields \citep{Li98,Va11}.
The overall MHD stability of these regions---including their propensity
to produce episodic bursts of solar wind \citep{Sh97,Wa00,SN04}---is
likely to depend sensitively on the physics of waves and their
dissipation as studied here.

\acknowledgments

The authors gratefully acknowledge
Richard Frazin and Alberto V\'{a}squez
for many valuable discussions.
This work was supported by NSF SHINE program grant AGS-1540094
and the University of Colorado's George Ellery Hale
Graduate Student Fellowship.

\end{document}